\documentclass[10pt]{article}
\usepackage{soul,color}
\usepackage{multicol}
\usepackage{graphicx}
\usepackage{amsmath}
\usepackage[a4paper]{geometry}
\usepackage{hyperref}

\begin{document}

\begin{center}
{\Large \bf Multi-source thermal model describing multi-region
structure of transverse momentum spectra of identified particles
and parameter dynamics of system evolution in relativistic
collisions}

\vskip1.0cm

Jia-Yu Chen$^{1,}${\footnote{E-mail:
202012602001@email.sxu.edu.cn}}, Mai-Ying
Duan$^{1,}${\footnote{E-mail: duanmaiying@sxu.edu.cn}}, Fu-Hu
Liu$^{1,}${\footnote{Corresponding author. E-mail:
fuhuliu@163.com; fuhuliu@sxu.edu.cn}}, Khusniddin K.
Olimov$^{2,3,}${\footnote{Corresponding author. E-mail:
khkolimov@gmail.com; kh.olimov@uzsci.net}}

{\small\it $^1$Institute of Theoretical Physics, State Key
Laboratory of Quantum Optics and Quantum Optics Devices \& \\
Collaborative Innovation Center of Extreme Optics, Shanxi
University, Taiyuan 030006, China

$^2$Laboratory of High Energy Physics, Physical-Technical
Institute of Uzbekistan Academy of Sciences, \\ Chingiz Aytmatov
Str. 2b, Tashkent 100084, Uzbekistan

$^3$Department of Natural Sciences, National University of Science
and Technology MISIS (NUST MISIS), \\ Almalyk Branch, Almalyk
110105, Uzbekistan}

\end{center}

\vskip1.0cm

{\bf Abstract:} In this article, the multi-region structure of
transverse momentum ($p_T$) spectra of identified particles
produced in relativistic collisions is studied by the
multi-component standard distribution (the Boltzmann, Fermi-Dirac,
or Bose-Einstein distribution) in the framework of a multi-source
thermal model. Results are interpreted in the framework of string
model phenomenology in which the multi-region of $p_T$ spectra
corresponds to the string hadronization in the cascade process of
string breaking. The contributions of the string hadronizations
from the first-, second-, and third- i.e. last-generations of
string breakings mainly form high-, intermediate-, and low-\(p_T\)
regions, respectively. From the high- to low-\(p_T\) regions, the
extracted volume parameter increases rapidly, and temperature and
flow velocity parameters decrease gradually. The multi-region of
\(p_T\) spectra reflects the volume, temperature, and flow
velocity dynamics of the system evolution. Due to the successful
application of the multi-component standard distribution, this
work reflects that the simple classical theory can still play a
great role in the field of complex relativistic collisions.
\\

{\bf Keywords:} transverse momentum spectra; multi-region
structure; multi-source thermal model; volume and temperature
dynamics; system evolution

{\bf PACS:} 12.40.Ee, 13.85.Hd, 24.10.Pa

\vskip1.0cm
\begin{multicols}{2}

{\section{Introduction}}

As physical quantities that can be measured in the experiments of
relativistic hadron-hadron, hadron-nucleus, and nucleus-nucleus
collisions, transverse momentum (\(p_T\)) spectra and other
abundant data of identified hadrons can shed light on the
mechanisms of particle production and characteristics of system
evolution~\cite{1,3,2,2a}. In particular, the \(p_T\) spectra
reflect the excitation degree of the system and the collective
motion of produced particles. The excitation degree can be
described by the thermal motion of particles, i.e. the temperature
of the system or source. The collective motion can be described by
the transverse flow velocity of particles or the transverse
velocity of source's expansion, which causes the spectra to be
blue-shifted. Generally, the contributions of thermal motion and
flow effect are intertwined. If the temperature parameter is
extracted from the data without excluding the contribution of flow
effect, we call it the effective temperature (\(T\)) which is not
an intrinsic (or thermal) temperature. Generally, the kinetic
freeze-out temperature (\(T_0\)) is the thermal temperature at the
kinetic freeze-out stage of the system evolution, in which the
flow effect is excluded in some way.

There are different methods to separate the contributions of
thermal motion and flow effect. These methods include, but are not
limited to, i) the method of blast-wave
model~\cite{3a,3b,3c,3d,3e} in which \(T_0\) and average
transverse flow velocity (\(\langle\beta_T\rangle\)) can be
obtained from the experimental \(p_T\) spectra by using a
probability density function in theory, ii) the method of applying
the Lorentz-like transformation of \(p_T\) in theoretical \(p_T\)
distribution~\cite{3f,3g,3h,3i} so that the function of \(T\) is
transformed into the function of \(T_0\) and
\(\langle\beta_T\rangle\), and iii) the alternative method of
intercept-slope~\cite{3j,3k,3l,3m,3p} in which \(T_0\) is obtained
from the intercept of the linear relation of \(T\) versus \(m_0\),
and \(\langle\beta_T\rangle\) is obtained from the slope of the
linear relation of \(\langle p_T\rangle\) versus \(\langle
m\rangle\), where \(\langle p_T\rangle\) is the average \(p_T\)
and \(\langle m\rangle\) is the average energy (the average mass
of the moving particles in the source rest frame). In the linear
relation of \(T\) versus \(m_0\), \(\langle\beta_T\rangle\) is
related to, but not equal to, the slope~\cite{3r,3s}.

The method of extracting \(T_0\) and \(\langle\beta_T\rangle\) is
not the main focus of this article. Instead, we focus our main
attention on the general structure, i.e. the various
regions~\cite{12-a,12-b,12-c}, of \(p_T\) spectra. As we know, the
\(p_T\) spectra cannot be fitted by a single function due to the
emissions of multiple sources according to the multi-source
thermal model~\cite{12,13,14,15}. Naturally and at first, one may
think that the \(p_T\) spectra can be divided into two regions:
the high- and low-\(p_T\) regions. One may use different functions
or distributions to fit the spectra in different \(p_T\) regions.
For example, one may use the two-component function, the Hagedorn
function or inverse power law from the quantum chromodynamics
(QCD) calculus~\cite{16a,16b,16c,16d} as the second component to
fit the spectra in the high-\(p_T\) region, and the standard
distribution (the Boltzmann, Fermi-Dirac, or Bose-Einstein
distribution) from the Boltzmann-Gibbs statistics or the Tsallis
distribution from the Tsallis statistics~\cite{16g,16h,16i,16j} as
the first component to fit the spectra in the low-\(p_T\) region
due to different scenarios in particle production.

In the fit of two-component function, the spectra in the
high-\(p_T\) region is regarded as the result of hard scattering
process which drives the increase of the particle yield in high
temperature region, and the spectra in the low-\(p_T\) region is
considered as the result of soft excitation process which
contributes to the low temperature region. If the two-component
function fails to fit the \(p_T\) spectra, one may consider the
three-component function in which the third, second, and first
components correspond to the high-, intermediate-, and low-\(p_T\)
regions, respectively, and the temperatures from the third,
second, and first components reduce orderly. Indeed, sometimes,
the \(p_T\) spectra have to be divided into three regions: the
high-, intermediate-, and low-\(p_T\) regions. The multi-component
function corresponds to the multi-region of \(p_T\) spectra, which
is a natural result of the multi-source thermal
model~\cite{12,13,14,15}.

As a statistical model, the multi-source thermal
model~\cite{12,13,14,15} does not provide the dynamic information
of the system evolution due to the fact that the time order of
particle production from different sources is not available in it,
though various temperatures corresponding to different \(p_T\)
regions can be obtained by using the same standard distribution or
Tsallis statistics~\cite{16g,16h,16i,16j}. The idea that the same
distribution can be applied in different \(p_T\) regions is based
on the similarity and universality that existed in relativistic
collisions~\cite{4,5,6,7,8,9,10,11}. It is necessary to note that
the multi-source thermal model~\cite{12,13,14,15} is combined with
other models such as the string model~\cite{16k,16l,16m}, and more
abundant information related to the system evolution can be
extracted from the \(p_T\) spectra.

In this article, we will study the method for describing the
\(p_T\) spectra in the multi-source thermal model. Combining with
the string model phenomenology, the idea and method for extracting
the effective temperatures of various sources at different states
of the system evolution are presented. As an application of the
simple idea and method, we will study the \(p_T\) spectra in
collisions at the Relativistic Heavy Ion Collider (RHIC) as an
example. Naturally, the \(p_T\) spectra in collisions at other
energies can be studied by the same idea and method. Although the
Tsallis distribution can be widely used in the description of the
$p_T$ spectra, it is not suitable to describe the multi-region
structure of the $p_T$ spectra due to the fact that it trowels the
structure. So, only the standard case will be studied by us to
describe the multi-region structure.

The remainder of this article is structured as follows. The
picture and formalism of the multi-source thermal model are
described in Section 2. An application of multi-component standard
distribution is given in Section 3. In Section 4, we give our
summary and conclusions.
\\

{\section{Formalism and method}}

According to the multi-source thermal model~\cite{12,13,14,15},
some emission sources are assumed to be formed in relativistic
collisions. In the model, there are different particle-products
which are originated from different kinds of sources, though the
same or similar functions or distributions can be used to describe
various particle-products and their sources. To understand the
whole picture of the multi-source in detail, let us first
introduce different sources of nuclear fragments and produced
particles. The picture of multi-source results in the multi-region
structure of transverse momentum spectra of nuclear fragments and
produced particles, though the formation mechanisms of emission
sources corresponding to the two kinds of products are different.

For nuclear fragments (such as a proton, neutron, deuteron,
triton, helium, lithium, etc.) emitted mainly from the spectator
region which is beyond the overlap area of projectile and target
nuclei, the emission sources may be nucleons and nucleon clusters.
For particles (such as \(\pi^{\pm}\), \(K^{\pm}\), and \(p\) which
are light flavor particles, and \(J/\psi\), \(\psi(2S)\), and
\(\it\Upsilon(\rm 1\it S)\) which are heavy flavor particles)
produced mainly in the participant zone that is in the overlap
area, the emission sources may be quarks or partons. A few
particles are produced in the spectator due to the cascade
collisions between the particles produced in the participant zone
and the nucleons of the spectator prefragment. Here, the
spectators are the fragments/free nucleons that are produced after
the deexcitation of the spectator prefragment. The number of
particles produced in such cascade decay depends on the collision
energy and size of participant/spectator zone. The emission
sources of the particles produced in the spectator may be nucleons
and/or other particles, which is similar to the particle
production in statistical bootstrap model of hadronic
matter~\cite{52-a,52-b,52-c}.

Because of the relative motion, there is a friction between the
spectator and participant. The contact layer between the spectator
and participant may get more heat to stay in a high degree of
excitation. The other part in the spectator may get less heat to
stay in a low degree of excitation. The contact layer is a hot
source, and the other part is a cold source, of nuclear fragments.
That is to say that, the nuclear fragments are emitted with
two-temperature. Considering that the transfer of friction heat
takes time, the nuclear fragments from the hot source are emitted
earlier than those from the cold source.

Not only for the cold source, but also for the hot source, one may
use the standard or Tsallis distribution to describe the \(p_T\)
spectra of nuclear fragments. The total result is the sum of the
contributions of two components (the cold or first component and
the hot or second component) with different temperatures (\(T_1\)
and \(T_2\)) and fractions (\(k\) and \(1-k\)), where \(k\) is the
fraction of cold reservoir. Generally, \(T_1<T_2\) and \(k>1-k\),
however, the relative size of \(k\) and \(1-k\) is not absolute.
If the spectator is small enough, the transfer time of friction
heat is negligible, we have \(T_1\approx T_2\). Thus, the
spectator fragments are simultaneously emitted with a single
temperature in case of a small spectator prefragment.

For particles produced in violent collisions in the participant
region, the temperatures of various sources are different due to
different timescales of source formations and particle
productions. The particles in the high-\(p_T\) region are produced
earlier than those in the low-\(p_T\) region. The source
temperature of particles in the high-\(p_T\) region is higher than
that in the low-\(p_T\) region. That is, we may use the standard
or Tsallis distribution of two-component (two-temperature) to
describe the \(p_T\) spectra of particles. The first component
with a low temperature \(T_1\) and a fraction \(k\) contributes in
the low-\(p_T\) region, and the second component with a high
temperature \(T_2\) and a fraction \(1-k\) contributes in the
high-\(p_T\) region. We have \(T_1<T_2\) if the particles in the
high-\(p_T\) region are produced earlier than those in the
low-\(p_T\) region, or \(T_1\approx T_2\) if the particles in both
the high- and low-\(p_T\) regions are produced at the nearly same
time.

In terms of string model phenomenology~\cite{16k,16l,16m}, many
strings are formed due to parton interactions in relativistic
collisions. These strings will undergo the processes of
hadronization and breaking. The particles in the high-\(p_T\)
region are produced in the hadronization process from the
first-generation string. If the first-generation string does not
undergo the hadronization process, it will undergo the breaking
process and more second-generation strings are formed. In the
two-component distribution, the particles in the low-\(p_T\)
region are produced in the hadronization process from the
second-generation string.

In the case of using the three-component distribution, the first
component describes the particles in the low-\(p_T\) region which
is contributed by the source with a low temperature \(T_1\) and a
fraction \(k_1\) from the hadronization of the third-generation
string. The second (third) component describes the particles in
the intermediate-\(p_T\) (high-\(p_T\)) region which is
contributed by the source with an intermediate (high) temperature
\(T_2\) (\(T_3\)) and a fraction \(k_2\) (\(k_3=1-k_1-k_2 \)) from
the hadronization of the second-generation (first-generation)
string, where the hadronization of the first-generation string
contributes in the high-\(p_T\) region. Generally, \(T_1<T_2<T_3\)
and \(k_1>k_2>k_3=1-k_1-k_2\).

Although the multi-source thermal model is a static
thermodynamical and statistical model, the dynamical evolution of
source temperature can be described by the model in terms of the
above idea from the \(p_T\) spectra. Comparing with the data in
different experimental conditions, the dependence of temperature
with collision energy, event centrality, nuclear size, particle
rapidity, particle mass, and quark mass can be obtained. The
behavior of the temperature dependence can be used to study the
mechanisms of particle production and characteristics of system
evolution.

Let us examine the \(i\)-th component in the multi-component
distribution. To study the \(p_T\) spectra from the most classic
theoretical consideration, we only use here the standard
distribution in the formalism representation. According to
ref.~\cite{3e}, one has the invariant yield of the particle
momentum (\(p\)) distribution to be
\begin{align}
E\frac{d^3N}{d^3p}\bigg|_i=&\ \frac{1}{2\pi
p_T}\frac{d^2N}{dydp_T}\bigg|_i \nonumber\\
=&\ \frac{gV_i}{(2\pi)^3}E
\bigg[\exp\bigg(\frac{E-\mu}{T_i}\bigg)+S\bigg]^{-1},
\end{align}
where \(E=\sqrt{p^2+m_0^2}=\sqrt{p_T^2+m_0^2}\cosh y\) is the
energy, \(m_0\) is the rest mass, \(y\) is the rapidity,
\(g=2s+1\) is the spin degeneracy factor, \(s\) is the spin
quantum number, and \(\mu\) is the chemical potential of the
considered particle; \(N\) is the number of identified particles
and \(V_i\) is the volume (normalization constant) of the source
or system at the temperature \(T_i\) (free parameter); while
\(S=0\), 1, and \(-1\) correspond to the Boltzmann, Fermi-Dirac,
and Bose-Einstein distributions, respectively. The density
function of momenta is
\begin{align}
\frac{dN}{dp}\bigg|_i=\frac{gV_i}{(2\pi)^3}4\pi p^2
\bigg[\exp\bigg(\frac{\sqrt{p^2+m_0^2}-\mu}{T_i}\bigg)+S\bigg]^{-1}
\end{align}
which is normalized to \(N\), where \(\sqrt{p^2+m_0^2}\) replaces
\(E\) to show the dependence on momentum \(p\) clearly.

The unit-density function of \(p_T\) and the rapidity \(y\) is
\begin{align}
\frac{d^2N}{dydp_T}\bigg|_i=&\ \frac{gV_i}{(2\pi)^2}p_T
\sqrt{p_T^2+m_0^2}\cosh y \nonumber\\
&\times\bigg[\exp\bigg(\frac{\sqrt{p_T^2+m_0^2}\cosh y
-\mu}{T_i}\bigg)+S\bigg]^{-1},
\end{align}
where \(\sqrt{p_T^2+m_0^2}\cosh y\) replaces \(E\) to show \(p_T\)
and \(y\) clearly. The density function of \(p_T\) is
\begin{align}
\frac{dN}{dp_T}\bigg|_i=&\ \frac{gV_i}{(2\pi)^2}p_T
\sqrt{p_T^2+m_0^2}\int_{y_{\min}}^{y_{\max}}\cosh y \nonumber\\
&\times\bigg[\exp\bigg(\frac{\sqrt{p_T^2+m_0^2}\cosh y
-\mu}{T_i}\bigg)+S\bigg]^{-1}dy,
\end{align}
where \(y_{\min}\) and \(y_{\max}\) denote the minimum and maximum
\(y\) respectively, in the considered mid-\(y\) bin. The density
function of \(y\) is
\begin{align}
\frac{dN}{dy}\bigg|_i =&\ \frac{gV_i}{(2\pi)^2}\cosh y
\int_{0}^{p_{T\max}}p_T\sqrt{p_T^2+m_0^2} \nonumber\\
&\times \bigg[\exp\bigg(\frac{\sqrt{p_T^2+m_0^2}\cosh y
-\mu}{T_i}\bigg)+S\bigg]^{-1}dp_T,
\end{align}
where \(p_{T\max}\) denotes the maximum \(p_T\), which is infinity
in mathematics and large enough (e.g. $\sim$ 30--50 GeV/$c$ and
above) when we fit the real data. The distribution range of \(y\)
in Eq. (5) is infinity which is certainly beyond
\([y_{\min},y_{\max}]\) in Eq. (4).

Equation (5) is the distribution of \(y\) in the rest frame of the
source. In fact, the longitudinal motions of a series of sources
with different rapidities in the one-dimensional rapidity space
have to be considered. The rapidity of the considered particle in
the moving frame of the source is then revised to \(y-Y\), where
\(Y\) is the source rapidity. We have the new expression for the
distribution of \(y\) to be
\begin{align}
\frac{dN}{dy}\bigg|_i =&\ \frac{1}{Y_{\max}-Y_{\min}}
\frac{gV_i}{(2\pi)^2}\int_{Y_{\min}}^{Y_{\max}}\cosh(y-Y) \nonumber\\
&\times \int_{0}^{p_{T\max}}p_T\sqrt{p_T^2+m_0^2} \nonumber\\
&\times\!
\bigg[\!\exp\!\bigg(\!\frac{\sqrt{p_T^2\!+\!m_0^2}\cosh(y\!-\!Y)\!
-\!\mu}{T_i}\!\bigg)\!+\!S\bigg]^{-1}\!dp_TdY,
\end{align}
where \(Y_{\min}\) and \(Y_{\max}\) denote the range of \(Y\),
which is mainly related to collision energy, but not the order
\(i\). Generally, \(|Y_{\min}|\) and \(|Y_{\max}|\) increase
linearly with the increase of logarithmic collision
energy~\cite{14}.

Considering \(n\) components in total, we have the multi-component
distribution or function to be
\begin{align}
E\frac{d^3N}{d^3p}=\frac{1}{2\pi
p_T}\frac{d^2N}{dydp_T}=\sum_{i=1}^n E\frac{d^3N}{d^3p}\bigg|_i,
\end{align}
\begin{align}
\frac{dN}{dp}=\sum_{i=1}^n \frac{dN}{dp}\bigg|_i,
\end{align}
\begin{align}
\frac{d^2N}{dydp_T}=\sum_{i=1}^n \frac{d^2N}{dydp_T}\bigg|_i,
\end{align}
\begin{align}
\frac{dN}{dp_T}=\sum_{i=1}^n \frac{dN}{dp_T}\bigg|_i,
\end{align}
\begin{align}
\frac{dN}{dy}=\sum_{i=1}^n \frac{dN}{dy}\bigg|_i.
\end{align}
In terms of \(p_T\) density, as an example, the ratio of the
contribution (\(dN/dp_T|_i\)) of the \(i\)-th component to that
(\(dN/dp_T\)) of \(n\) components is defined as the fraction
\(k_i\) of the \(i\)-th component, where \(k_i\) are the
parameters that do not appear obviously in the above equations.
Generally, the normalization gives \(\sum_{i=1}^{n}k_i=1\). In the
fit of data, Eqs. (7)--(11) can be used in the right way according
to different styles of data set.

The above discussions are applicable for other distributions such
as the Tsallis distribution~\cite{16g,16h,16i,16j}. In fact, for
very large number of components, the standard case should result
in the Tsallis distribution which has the invariant yield of
particle distribution to be
\begin{align}
E\frac{d^3N}{d^3p}\bigg|_i=&\ \frac{1}{2\pi
p_T}\frac{d^2N}{dydp_T}\bigg|_i \nonumber\\
=&\ \frac{gV_i}{(2\pi)^3}E
\bigg[1+(q-1)\frac{E-\mu}{T_i}\bigg]^{-\frac{q}{q-1}},
\end{align}
where \(q\) is the entropy index which describes the degree of
non-equilibrium. Generally, \(q>1\), while \(q=1\) means an
equilibrium. A larger \(q\) corresponds to a further deviation
from the equilibrium. We have used the assumption of \(E>\mu\). If
\(E<\mu\), \(-E+\mu\) and \(q/(q-1)\) should replace \(E-\mu\) and
\(-q/(q-1)\) in the equation respectively.

The above discussions are also suitable for the Tsallis-standard
(Tsallis form of standard) distribution which describes the
invariant yield of particle distribution in two forms as
\begin{align}
E\frac{d^3N}{d^3p}\bigg|_i=&\ \frac{1}{2\pi
p_T}\frac{d^2N}{dydp_T}\bigg|_i \nonumber\\
=&\ \frac{gV_i}{(2\pi)^3}E
\bigg\{\bigg[1+(q-1)\frac{E-\mu}{T_i}\bigg]^{\frac{q}{q-1}}+S\bigg\}^{-1} \nonumber\\
{\rm or}& \ \ \frac{gV_i}{(2\pi)^3}E
\bigg\{\bigg[1+(q-1)\frac{E-\mu}{T_i}\bigg]^{\frac{1}{q-1}}+S\bigg\}^{-q}.
\end{align}
Here, we have used again the assumption of \(E>\mu\). If
\(E<\mu\), \(-E+\mu\) and \(-q/(q-1)\) should replace \(E-\mu\)
and \(q/(q-1)\) in the equation respectively.

Besides the Tsallis and Tsallis-standard distributions, the
\(q\)-dual and \(q\)-dual-standard distributions can also describe
the \(p_T\) spectra in the region of 10 GeV/$c$ and above.
According to ref.~\cite{52}, the \(q\)-dual-standard distribution
represents the invariant particle yield in the form of
\begin{align}
E\frac{d^3N}{d^3p}\bigg|_i=&\ \frac{1}{2\pi
p_T}\frac{d^2N}{dydp_T}\bigg|_i \nonumber\\
=&\ \frac{gV_i}{(2\pi)^3}E \sum_{n=0}^{\infty}(-S)^n \nonumber\\
&\times\bigg[1+(n+1)(q-1)\frac{E-\mu}{T_i}\bigg]^{-\frac{q}{q-1}},
\end{align}
where \(n\) is an integer greater than or equal to 0. In a real
fit, we can take the maximum value of \(n\) to be 10 based on our
confirmatory calculation in which the contributions of the terms
with $n>10$ can be neglected. If $S=0$ in the $q$-dual-standard
distribution, we write the \(q\)-dual distribution as
\begin{align}
E\frac{d^3N}{d^3p}\bigg|_i=&\ \frac{1}{2\pi
p_T}\frac{d^2N}{dydp_T}\bigg|_i \nonumber\\
=&\ \frac{gV_i}{(2\pi)^3}E \sum_{n=0}^{\infty}
\bigg[1+(n+1)(q-1)\frac{E-\mu}{T_i}\bigg]^{-\frac{q}{q-1}}.
\end{align}

In the above equations, although \(\mu\) can be regarded as a free
parameter, it is nearly zero at relativistic
energies~\cite{51-a,51-b,51-c,51-d}. Therefore, the assumption of
\(E>\mu\) can be satisfied in general. In the fit, the main
parameters are \(V_i\) and \(T_i\), while \(k_i=V_i/\sum V_i \)
that is expressed as a function of \(V_i\). Generally, at first,
we use the first component to fit the spectra as wide as possible
in low-\(p_T\) region. Then, we add the second component to fit
the spectra as wide as possible in intermediate-\(p_T\) region.
Finally, we add the third component to fit the spectra in
high-\(p_T\) region. It should be noted that when the second
component is included, the parameters of the first component
should be fixed to the value found before.

However, due to the influence of two sets of parameters from the
two components, the parameters of the first component will have
some changes to obtain the minimum $\chi^2$. In the fit, the
method of least squares is used to determine the best values of
parameters, and the method of statistical simulation~\cite{51a} is
used to determine the uncertainties of parameters in our work.
Compared with fitting three components simultaneously without
fixing any parameters in advance, the stepwise fitting method used
by us can give the results more quickly. Generally, the two
fitting methods can get consistent values of parameters because we
have obtained the minimum $\chi^2$ to determine various
parameters.

Our previous work~\cite{51} shows that the temperature parameter
extracted from the standard distribution is lower than the one
from the Tsallis distribution, and the temperature parameter
extracted from the Tsallis distribution is lower than the one from
the Tsallis-standard distribution. A two- or three-component
standard distribution can be covered by the Tsallis distribution,
and a two- or three-component Tsallis distribution can be covered
by the Tsallis-standard distribution. Because the standard
distribution is the most basic and classical distribution, we are
inclined to use the standard distribution with multi-component in
the fit of multi-region of \(p_T\) spectra. The multi-temperature
is a description of various stages of the system evolution.

In the fit of standard distribution, we need a two- or
three-component function. The fine structure of \(p_T\) spectra
can be described by the two- or three-component standard
distribution which results from the multi-source thermal
model~\cite{12,13,14,15}. Generally, the first component describes
the spectra in low-\(p_T\) region which results from the
hydrodynamic evolution of partons. The second component describes
the spectra in intermediate-\(p_T\) region which originates from
the coalescence and fragmentation of partons. Meanwhile, the third
component describes the spectra in high-\(p_T\) region which
derives from the hard-scattering of partons.

Because of the underlying participants or contributors being
partons for the spectra in whole \(p_T\) region, the maximum
energy density of partons determines common behavior and basic
property of \(p_T\) spectra. If the multi-component standard
distribution describes the fine structure of \(p_T\) spectra, the
single distributions such as the Tsallis, Tsallis-standard,
\(q\)-dual, and \(q\)-dual-standard distributions do not reveal
the fine structure in wide \(p_T\) region. Although the mentioned
single distributions with fewer parameters are more suitable for
the fit, the multi-component standard distribution can be used to
extract the volume and temperature dynamics of the system
evolution.

Before fitting the $p_T$ spectra using the multi-component
standard distribution, we would like to point out that the
temperature fluctuation is a way to explain the origin of the
nonextensive distributions of the Tsallis statistics. Because of
the temperature fluctuation, the interactions among the different
local sources or subsystems happen through the exchange of heat
energy, and result in the couplings of entropy functions of
various subsystems. As a result, the total entropy is the sum of
entropies of subsystems plus entropies of the couplings.

These nonextensive distributions of the Tsallis statistics are
related to the Boltzmann's factor by a continuous summation of the
Boltzmann factor weighed by factors given by the Euler-Gamma
function. Thus, one can expect that, for proton-proton collisions,
adding a sufficiently high number of components would approximate
the result to a Tsallis distribution. For nucleus-nucleus
collisions, the distribution would be deformed, with the effects
of the collective flow modifying the Euler-Gamma function.

As mentioned above, in the multi-component distribution, different
components correspond to different mechanisms of particle
production. Generally, the feed-down contributions of resonances
are in the very low-\(p_T\) region~\cite{51b}, which is naturally
considered by us in the first component and contributed by the
third- i.e. the last-generation of string breakings. If the
relative contributions of resonances are large, a small $T_1$ and
a large $k_1$ will be obtained. On the contrary, a large $T_1$ and
a small $k_1$ will be obtained if the relative contributions of
resonances are small. Here, we emphasize that the first component
describes the spectra in low-\(p_T\) region which results from the
third-generation of string breakings.
\\

{\section{Results and discussion}}

\begin{figure*}[htb!]
\begin{center}
\includegraphics[width=10.0cm]{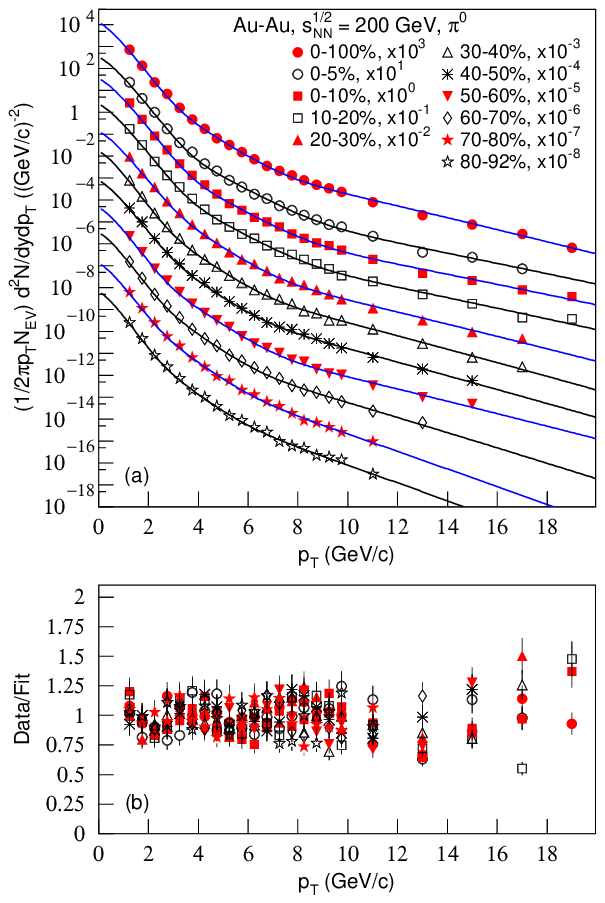}
\end{center}
Fig. 1. (a) The invariant yields, \((1/2\pi p_TN_{EV})
d^2N/dydp_T\), of \(\pi^0\) produced in \(|\eta|<0.35\) in Au-Au
collisions with different centralities at \(\sqrt{s_{NN}}=200\)
GeV. The symbols represent the experimental data measured by the
PHENIX Collaboration~\cite{55}. The curves are our fits by the
multi-component standard distribution, Eqs. (1) and (7). (b)
Values of Data/Fit corresponding to the fits in panel (a).
\end{figure*}

To study the temperature dynamics of the system evolution, we use
the multi-component standard distribution in the framework of
multi-source thermal model to analyze the spectra of various
particles at different energies. As an example, we analyze the
spectra of neutral pions (\(\pi^0\)) produced in collisions at the
RHIC. Figure 1(a) shows the invariant yields, \((1/2\pi p_TN_{EV})
d^2N/dydp_T\), of \(\pi^0\) produced in mid-pseudorapidity
(\(|\eta|<0.35\)) in gold-gold (Au-Au) collisions in different
centralities at the center-of-mass energy per nucleon pair
\(\sqrt{s_{NN}}=200\) GeV, where \(N_{EV}\) denotes the number of
events. The various symbols represent the experimental data
measured by the PHENIX Collaboration~\cite{55}. The curves are our
fits with the multi-component standard distribution. The samples
for different centralities are re-scaled by different factors
marked in the panel. Following Figure 1(a), the values of Data/Fit
corresponding to the fits are shown in Figure 1(b). The values of
normalization constants $V_1$, $V_2$, and $V_3$, free parameters
$T_1$, $T_2$, and $T_3$, as well as $\chi^{2}$ and number of
degrees-of-freedom (ndof) are listed in Table 1. One can see that
the PHENIX data are well fitted by the multi-component standard
distribution.

\begin{table*}[htp!]
\vskip1.0cm {\small Table 1. Values of normalization constants
$V_1$, $V_2$, and $V_3$, free parameters $T_1$, $T_2$, and $T_3$,
as well as $\chi^{2}$ and ndof corresponding to the solid curves
in Figures 1--3, where the values of $V_1$, $V_2$, and $V_3$ for
Figure 3 are obtained from the re-normalization from cross-section
to yield so that we may compare them with those for Figures 1 and
2 uniformly.
\begin{center}
\vspace{-0.2cm} \tiny
\begin{tabular} {ccccccccc}\\ \hline
Figure & Selection & $V_1$ (fm$^3$) & $V_2$ (fm$^3$) & $V_3$ (fm$^3$) & $T_1$ (MeV) & $T_2$ (MeV) & $T_3$ (MeV) & $\chi^2$/ndof \\
\hline
Figure 1 & 0--100\%      & $15179.314\pm1609.007$ & $20.427\pm1.626$ & $0.005\pm0.001$ & $257.3\pm3.3$ & $545.5\pm4.8$ & $1440.0\pm9.4$  & 18/17\\
Au-Au    & 0--5\%        & $37593.234\pm4586.375$ & $12.253\pm1.225$ & $0.003\pm0.001$ & $270.1\pm3.7$ & $644.7\pm6.3$ & $1662.8\pm23.1$ & 25/16\\
200 GeV  & 0--10\%       & $39159.617\pm4620.834$ & $12.898\pm1.187$ & $0.003\pm0.001$ & $270.1\pm3.4$ & $632.1\pm5.6$ & $1679.6\pm19.3$ & 28/17\\
         & 10--20\%      & $26932.924\pm4228.469$ & $12.275\pm1.522$ & $0.004\pm0.001$ & $270.1\pm4.6$ & $625.8\pm8.1$ & $1614.2\pm24.2$ & 47/17\\
         & 20--30\%      & $15241.632\pm1950.929$ & $32.015\pm3.130$ & $0.008\pm0.001$ & $267.8\pm4.0$ & $540.0\pm5.5$ & $1405.5\pm15.6$ & 25/16\\
         & 30--40\%      & $21424.977\pm2067.510$ & $38.813\pm3.027$ & $0.015\pm0.001$ & $245.0\pm3.7$ & $514.1\pm4.1$ & $1287.1\pm9.5$  & 17/16\\
         & 40--50\%      &  $9237.099\pm715.875$  & $25.366\pm1.700$ & $0.011\pm0.001$ & $257.8\pm2.3$ & $514.1\pm3.2$ & $1261.9\pm8.7$  & 13/15\\
         & 50--60\%      &  $5590.671\pm768.158$  & $13.275\pm1.147$ & $0.003\pm0.001$ & $247.0\pm4.2$ & $523.5\pm4.9$ & $1391.6\pm14.6$ & 25/15\\
         & 60--70\%      &  $4166.770\pm475.012$  & $15.232\pm1.226$ & $0.008\pm0.001$ & $241.8\pm3.4$ & $487.9\pm4.2$ & $1150.6\pm10.0$ & 13/14\\
         & 70--80\%      &  $1619.227\pm182.973$  &  $8.987\pm0.681$ & $0.011\pm0.001$ & $243.9\pm3.4$ & $478.3\pm4.1$ &  $999.4\pm7.1$  & 13/13\\
         & 80--92\%      &   $847.819\pm92.412$   &  $3.985\pm0.327$ & $0.005\pm0.001$ & $234.0\pm3.2$ & $464.0\pm3.8$ &  $979.4\pm8.1$  & 16/13\\
Figure 2 & 200 GeV       &$19489.150\pm2436.144$& $23.238\pm3.044$ & $0.010\pm0.001$ & $255.0\pm5.1$ & $551.0\pm7.2$ & $1357.3\pm12.2$ & 29/18\\
Cu-Cu 0--10\% & 62.4 GeV &$16881.004\pm2886.652$& $16.101\pm2.505$ & $-$             & $234.3\pm5.2$ & $475.0\pm6.7$ & $-$             & 36/8\\
Figure 3 & 200 GeV       &   $239.849\pm52.287$ &  $0.128\pm0.026$ & $0.000022\pm0.000004$ & $211.0\pm8.0$ & $534.9\pm9.5$ & $1398.2\pm28.0$ & 85/19\\
$pp$     & 62.4 GeV      &   $204.263\pm59.233$ &  $0.030\pm0.006$ & $-$                   & $199.9\pm7.8$ & $492.7\pm8.8$ & $-$             & 108/8\\
\hline
\end{tabular}%
\end{center}}
\end{table*}

\begin{figure*}[htb!]
\begin{center}
\includegraphics[width=10.0cm]{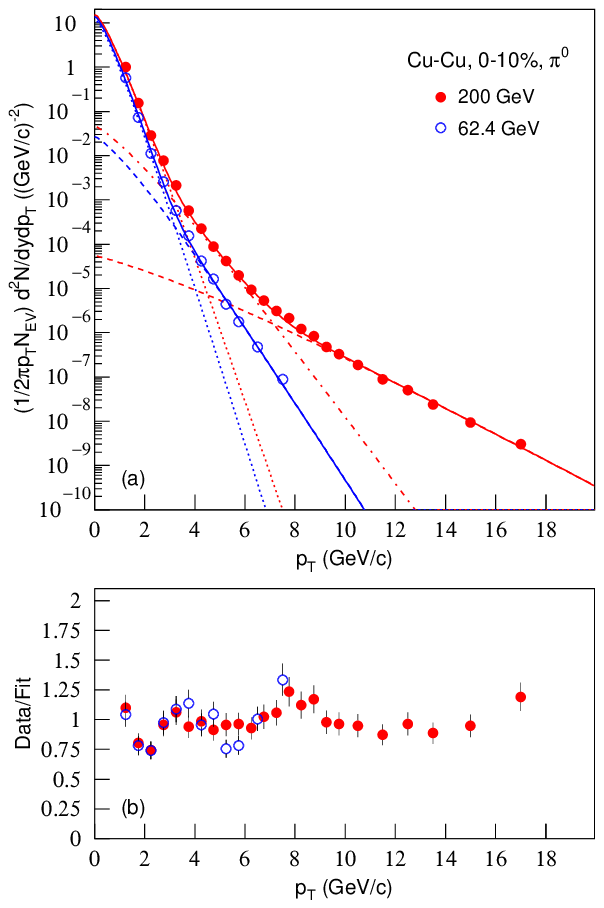}
\end{center}
Fig. 2. (a) The invariant yields, \((1/2\pi p_TN_{EV})
d^2N/dydp_T\), of \(\pi^0\) produced in \(|\eta|<0.35\) in 0--10\%
Cu-Cu collisions at \(\sqrt{s_{NN}}=200\) and 62.4 GeV. The
symbols represent the experimental data measured by the PHENIX
Collaboration~\cite{56}. The curves are our fits by the
multi-component standard distribution, Eqs. (1) and (7). (b)
Values of Data/Fit corresponding to the fits in panel (a).
\end{figure*}

\begin{figure*}[htb!]
\begin{center}
\includegraphics[width=10.0cm]{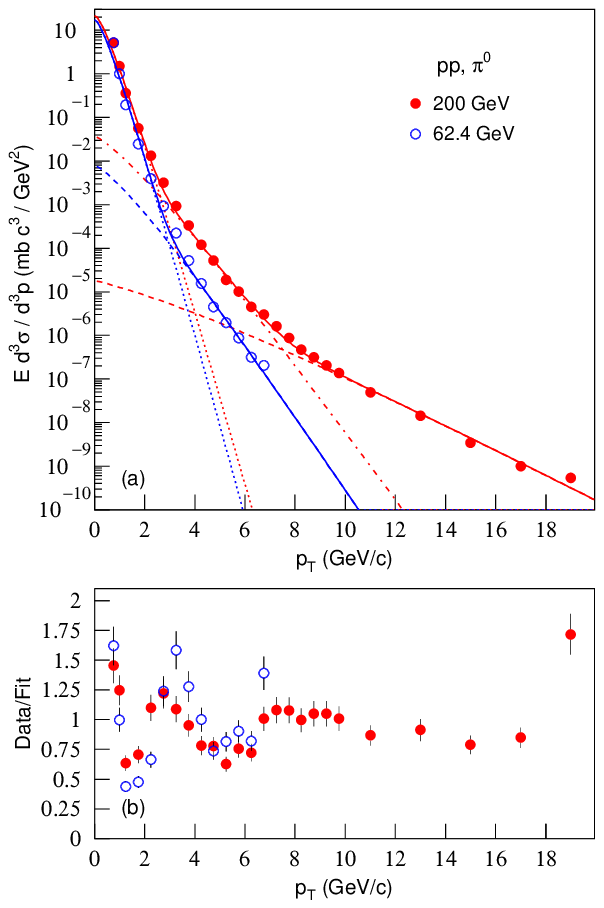}
\end{center}
Fig. 3. (a) The invariant cross-sections, \(Ed^3\sigma/d^3p\), of
\(\pi^0\) produced in \(|\eta|<0.35\) in \(pp\) collisions at
\(\sqrt{s_{NN}}=200\) and 62.4 GeV. The symbols represent the
experimental data measured by the PHENIX Collaboration~\cite{56}.
The curves are our fits by the multi-component standard
distribution, Eqs. (1) and (7). (b) Values of Data/Fit
corresponding to the fits in panel (a).
\end{figure*}

Figure 2(a) shows the invariant yields, \((1/2\pi p_TN_{EV})
d^2N/dydp_T\), of \(\pi^0\) produced in \(|\eta|<0.35\) in 0--10\%
copper-copper (Cu-Cu) collisions at \(\sqrt{s_{NN}}=200\) and 62.4
GeV. In addition, Figure 3(a) shows the invariant cross-sections,
\(Ed^3\sigma/d^3p\), of \(\pi^0\) produced in \(|\eta|<0.35\) in
proton-proton (\(pp\)) collisions at \(\sqrt{s_{NN}}=200\) and
62.4 GeV, where \(\sigma\) denotes the cross-section of \(\pi^0\)
production. The symbols represent the experimental data measured
by the PHENIX Collaboration~\cite{56}. The solid curves are our
fits by the multi-component standard distribution. At 200 GeV,
three-component function provides better results. The
contributions of the first, second, and third components
corresponding to the third, second, and first generations of
string breakings are represented by the dotted, dot-dashed, and
dashed curves, respectively, to underline their particular roles
in the formation of the spectra. At 62.4 GeV, the two-component
function proves suitable. The roles of the first and second
components are illustrated by the dotted and dashed curves,
accordingly. Following Figures 2(a) and 3(a), the values of
Data/Fit corresponding to the fits are shown in Figures 2(b) and
3(b) respectively. The values of parameters, $\chi^{2}$, and ndof
are listed in Table 1, where the values of $V_1$, $V_2$, and $V_3$
for Figure 3 are obtained from the re-normalization from the
cross-section to yield so that we may compare these parameters
uniformly. One can see again that the PHENIX data are well-fitted
by the multi-component standard distribution.

Although the Data/Fit ratios oscillate around the unity, this can
be attributed to the statistics of the experimental data. In
addition, there are structures in the ratio plots, with groups of
adjacent points deviating from the unity in one or another
direction. There is a special oscillation for Cu-Cu data around
the transverse momentum of 8 GeV/$c$, which is the place where the
dominating component of the fit changes. Larger oscillation for
\(pp\) data is caused by the small interaction volume which does
not result in more equilibrium than large interaction volume. In
our opinion, this behavior shows that the model describes
approximately the data due to the fact that it is formally
consistent with the experimental errors. In the end, the
$\chi^2$/ndof values, on which the experimental errors of the data
have a strong influence, are acceptable due to small
$\chi^2$/ndof, though the fit quality in a few cases in Figures 2
and 3 is not very satisfactory due to high values of
$\chi^2$/ndof. In general, the present fits are acceptable, though
four-component function with 8 parameters is expected to possibly
obtain a better fit.

From Figures 1(b)--3(b) and Table 1, one can see that the fitting
result in nucleus-nucleus collisions is better than that in \(pp\)
collisions. This is explained by the large volume and then more
equilibrium in nucleus-nucleus collisions. In the system with
large volume, the thermodynamic equilibrium is easier to achieve.
As a result, the standard distribution is more applicable.

From Table 1 one can see the tendencies of parameters. To see
clearly, the dependences of \(V_1\), \(V_2\), and \(V_3\) (upper
panel), as well as \(T_1\), \(T_2\), and \(T_3\) (lower panel), on
the centrality \(C\) in 200 GeV Au-Au collisions are given in
Figure 4. Meanwhile, the dependences of the free fitting
parameters in different centrality classes (\(C)\) relative to the
peripheral centrality class (80--92\%),
\((V_i-V_i|_{80-92\%})/V_i|_{80-92\%}\) (upper panel) and
\((T_i-T_i|_{80-92\%})/T_i|_{80-92\%}\) (lower panel), on \(C\) in
200 GeV Au-Au collisions are displayed in Figure 5. As marked in
the panels, the different symbols represent different parameters.
One can see that with the decrease of centrality from central to
peripheral collisions, \(V_1\) decreases significantly, \(V_2\)
has an increasing trend from central to semi-central collisions
and a decreasing trend from semi-central to peripheral collisions,
and \(V_3\) fluctuates somehow and has no obvious increasing or
decreasing trend. Meanwhile, with the decrease of centrality,
\(T_1\) decreases slightly, \(T_3\) decreases significantly, and
the decreasing degree of \(T_2\) is between \(T_1\) and \(T_3\).

\begin{figure*}[htb!]
\hspace{0.75cm}
\begin{center}
\includegraphics[width=10.0cm]{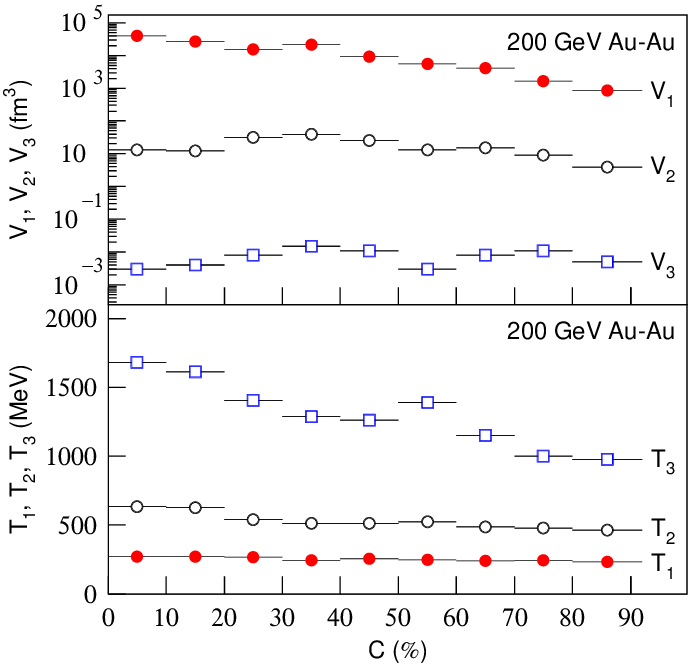}
\end{center}
Fig. 4. Dependences of normalization constants $V_1$, $V_2$, and
$V_3$ (upper panel), as well as free parameters $T_1$, $T_2$, and
$T_3$ (lower panel), on the centrality \(C\) in 200 GeV Au-Au
collisions. Different symbols represent different parameters
marked in the panels.
\end{figure*}

\begin{figure*}[htb!]
\begin{center}
\includegraphics[width=10.0cm]{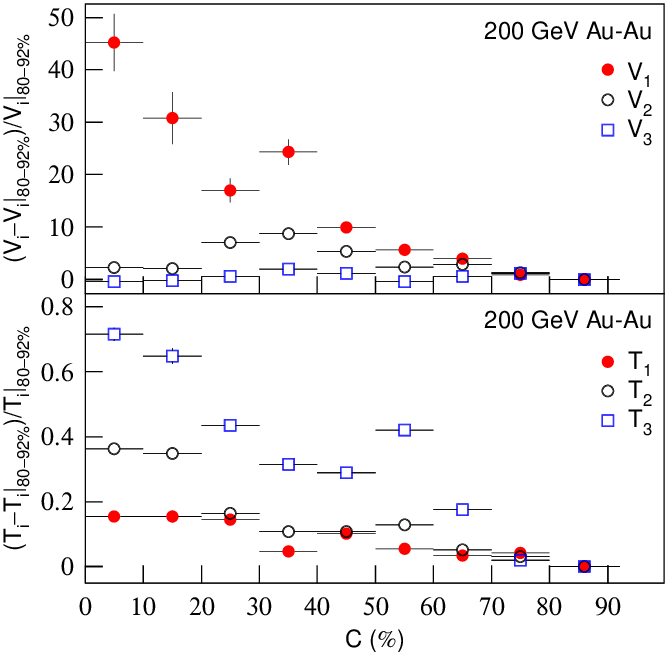}
\end{center}
Fig. 5. Dependences of \((V_i-V_i|_{80-92\%})/V_i|_{90-92\%}\)
(upper panel) and \((T_i-T_i|_{80-92\%})/T_i|_{90-92\%}\) (lower
panel) on \(C\) in 200 GeV Au-Au collisions. Different symbols
represent different parameters marked in the panels.
\end{figure*}

At the same energy per nucleon pair, the parameters \(V_i\)
decrease significantly and the parameters \(T_i\) decrease
slightly from nucleus-nucleus to \(pp\) collisions in most cases.
Although an increase appears from Au-Au to Cu-Cu collisions for
\(V_2\), it does not affect obviously the tendency of volume
parameter which is mainly determined by \(V_1\). Generally,
\(V_1\) and \(T_1\), which determine the volume and temperature of
the system respectively, decrease with the decrease of nuclear
size.

From the first generation strings to the third ones, the collision
system expands rapidly from the volume \(V_3\) to \(V_2\) and then
to \(V_1\). It is expected that the number of the third generation
strings is very large, which results in very large volume at the
final kinetic freeze-out stage. We believe that the first
generation strings are produced earlier than the second ones, and
the second ones are produced earlier than the third ones. This
renders the volume order of \(V_3<V_2<V_1\) and the temperature
order of \(T_3>T_2>T_1\), which reflects the volume and
temperature dynamics of the system evolution.

At 62.4 GeV, only two generation strings are needed and smaller
parameters are obtained. It is expected that single generation
strings are suitable to fit the \(p_T\) spectra at lower energy,
though the single generation strings may exist for a considerable
time. At higher energy, it is expected that four or more
generation strings can appear. Generally, in relativistic
collisions, the string breaking and hadronization depend on energy
of the collision, no matter how fast the projectile and target
passes through each other.

It should be noted that the temperature parameters \(T_i\)
extracted from this work contains the influence of collective flow
which results in \(T_i\) being larger than the kinetic freeze-out
temperature \(T_0\). As discussed in Section 1, one has a few
methods to extract \(T_0\) and the transverse flow velocity
\(\langle\beta_T\rangle\). Based on these methods, there are many
works~\cite{3a,3b,3c,3d,3i,3j,3k,3l} which studied the extractions
of \(T_0\) and \(\langle\beta_T\rangle\). Although the main focus
of this work is on a simple and useful method for studying the
fine multi-region structure of the \(p_T\) spectra, we may use an
almost model-independent approach to extract \(T_0\) and
\(\langle\beta_T\rangle\).

According to ref.~\cite{65}, one has the relation, $\langle
p_T\rangle=3.07T_0$. Then, we have the almost model-independent
$T_0=\langle p_T\rangle/3.07$ and
$\langle\beta_T\rangle=(2.07/3.07)\langle p_T\rangle/\langle
m\rangle$. One can see that $T_0$ and $\langle\beta_T\rangle$
depends mainly on $\langle p_T\rangle$ and $\langle m\rangle$
which originate from the data, but not from the model. However,
the factor 3.07 in this approach comes from the model~\cite{65}.
Therefore, we say this approach being the almost
model-independent. In the case of applying the string model
phenomenology, a string is formed by the interaction of two
partons via exchange of virtual gluons. For each parton, we should
use $\langle p_T\rangle/2$ instead of $\langle p_T\rangle$ in the
expressions of $T_0$ and $\langle\beta_T\rangle$. At the same
time, $\langle m\rangle$ is the constituent mass (0.31
GeV/$c$~\cite{66}) of up or down quarks multiplied by the average
Lorentz factor $\langle\gamma\rangle$, where
$\langle\gamma\rangle\approx6$ from the average energy~\cite{3p}
of pions in the rest frame of emission source at the RHIC. Here,
we have used the method for calculating the average energy of
particles in the source rest frame discussed in collisions at
higher energy~\cite{3p}.

Based on the discussion of the almost model-independent $T_0$ and
$\langle\beta_T\rangle$, our calculations show that $T_0$
approximately equals to 122--109 MeV and $\langle\beta_T\rangle$
approximately equals to 0.135--0.122 $c$ from central to
peripheral Au-Au collisions. Meanwhile, $T_0\approx118$ and 106
MeV, $\langle\beta_T\rangle\approx0.131$ and 0.118 $c$, for
central Cu-Cu collisions at 200 and 62.4 GeV respectively;
$T_0\approx97$ and 91 MeV, $\langle\beta_T\rangle\approx0.108$ and
0.101 $c$, for $pp$ collisions at 200 and 62.4 GeV respectively.
These results are based on a comprehensive analysis for three- or
two-component in the $p_T$ range from 0 to 30 GeV/$c$ and show
that $T_0$ and $\langle\beta_T\rangle$ decrease with the decrease
of centrality, nuclear size, and collision energy.

To study the $T_0$ and $\langle\beta_T\rangle$ dynamics of the
system evolution in detail, we now calculate $T_{0i}$ and
$\langle\beta_T\rangle_i$ which are the kinetic freeze-out
temperature and average transverse flow velocity of the $i$-th
component respectively. In the calculation of the $i$-th
component, the contributions of other components are excluded.
Table 2 shows the values of extracted parameters $T_{01}$,
$T_{02}$, $T_{03}$, $\langle\beta_T\rangle_1$,
$\langle\beta_T\rangle_2$, and $\langle\beta_T\rangle_3$
corresponding to the solid curves in Figures 1--3. To present the
tendencies of the kinetic freeze-out parameters on centrality
intuitively, these values are also displayed in Figures 6 and 7,
where the latter is in terms of relative values as Figure 5.

\begin{figure*}[htb!]
\hspace{0.75cm}
\begin{center}
\includegraphics[width=10.0cm]{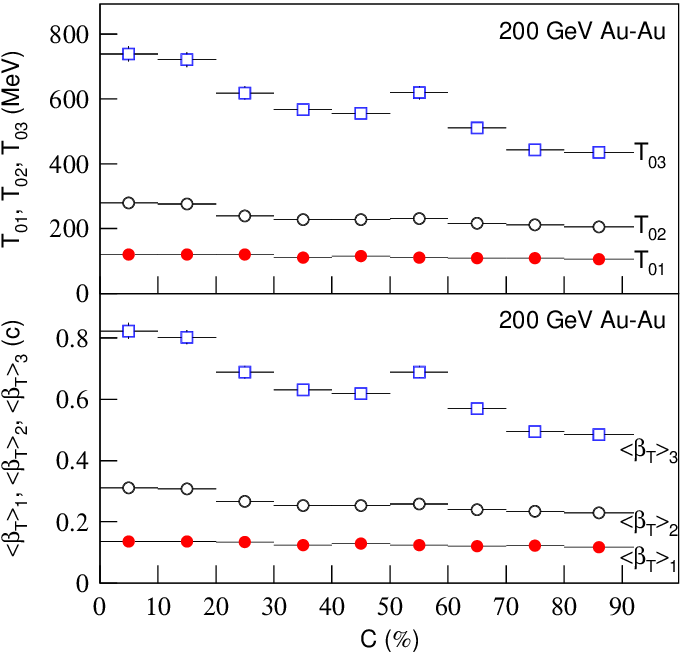}
\end{center}
Fig. 6. Dependences of extracted parameters $T_{01}$, $T_{02}$,
and $T_{03}$ (upper panel), as well as $\langle\beta\rangle_1$,
$\langle\beta\rangle_2$, and $\langle\beta\rangle_3$ (lower
panel), on the centrality \(C\) in 200 GeV Au-Au collisions.
Different symbols represent different parameters marked in the
panels.
\end{figure*}

\begin{figure*}[htb!]
\hspace{.0cm}
\begin{center}
\includegraphics[width=10.0cm]{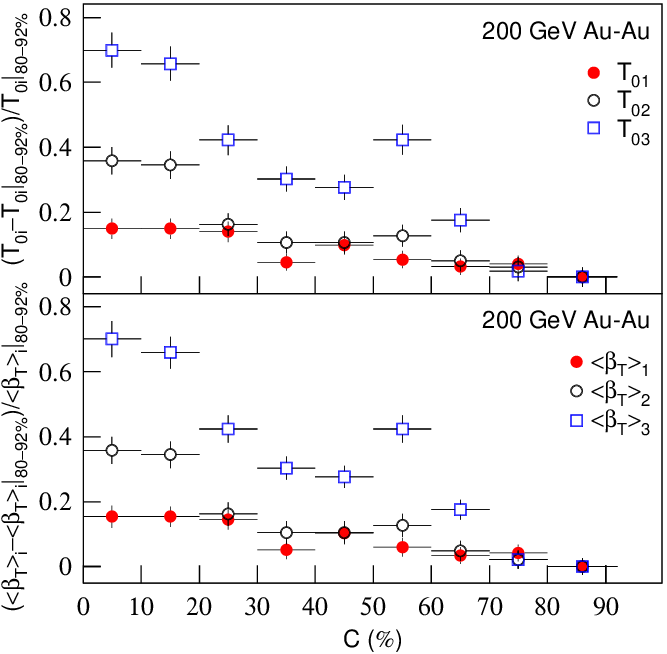}
\end{center}
Fig. 7. Dependences of
\((T_{0i}-T_{0i}|_{80-92\%})/T_{0i}|_{90-92\%}\) (upper panel) and
\((\langle\beta\rangle_i-\langle\beta\rangle_i|_{80-92\%})/\langle\beta\rangle_i|_{90-92\%}\)
(lower panel) on \(C\) in 200 GeV Au-Au collisions. Different
symbols represent different parameters marked in the panels.
\end{figure*}

\begin{table*}
{\small Table 2. Values of extracted parameters $T_{01}$,
$T_{02}$, $T_{03}$, $\langle\beta_T\rangle_1$,
$\langle\beta_T\rangle_2$, and $\langle\beta_T\rangle_3$
corresponding to the solid curves in Figures 1--3. \vspace{-.25cm}
\begin{center}
\small
\begin{tabular} {ccccccccc}\\ \hline
Figure & Selection & $T_{01}$ (MeV) & $T_{02}$ (MeV) & $T_{03}$
(MeV) & $\langle\beta_T\rangle_1$ ($c$) &
$\langle\beta_T\rangle_2$ ($c$)
& $\langle\beta_T\rangle_3$ ($c$) \\
\hline
Figure 1 & 0--100\%      & $115.4\pm3.0$ & $241.2\pm7.2$ & $638.1\pm21.2$ & $0.128\pm0.003$ & $0.268\pm0.008$ & $0.710\pm0.022$ \\
Au-Au    & 0--5\%        & $121.0\pm3.3$ & $284.7\pm8.8$ & $732.0\pm22.9$ & $0.135\pm0.003$ & $0.317\pm0.010$ & $0.815\pm0.025$ \\
200 GeV  & 0--10\%       & $121.0\pm3.3$ & $279.1\pm8.5$ & $739.3\pm23.4$ & $0.135\pm0.004$ & $0.311\pm0.010$ & $0.823\pm0.027$ \\
         & 10--20\%      & $121.0\pm3.3$ & $276.4\pm8.7$ & $721.2\pm22.6$ & $0.135\pm0.004$ & $0.308\pm0.009$ & $0.803\pm0.024$ \\
         & 20--30\%      & $120.0\pm3.2$ & $238.8\pm7.1$ & $618.9\pm20.0$ & $0.134\pm0.004$ & $0.266\pm0.008$ & $0.689\pm0.020$ \\
         & 30--40\%      & $110.1\pm2.7$ & $227.4\pm6.9$ & $566.8\pm16.8$ & $0.123\pm0.003$ & $0.253\pm0.008$ & $0.631\pm0.017$ \\
         & 40--50\%      & $115.7\pm3.0$ & $227.4\pm6.9$ & $555.7\pm16.6$ & $0.129\pm0.003$ & $0.253\pm0.008$ & $0.618\pm0.016$ \\
         & 50--60\%      & $111.0\pm2.7$ & $231.6\pm7.0$ & $619.2\pm20.0$ & $0.124\pm0.003$ & $0.258\pm0.008$ & $0.689\pm0.020$ \\
         & 60--70\%      & $108.7\pm2.7$ & $216.0\pm6.6$ & $511.3\pm16.0$ & $0.121\pm0.003$ & $0.240\pm0.007$ & $0.569\pm0.015$ \\
         & 70--80\%      & $109.6\pm2.7$ & $211.8\pm6.4$ & $443.5\pm13.4$ & $0.122\pm0.003$ & $0.234\pm0.006$ & $0.494\pm0.013$ \\
         & 80--92\%      & $105.3\pm2.6$ & $205.5\pm6.2$ & $435.2\pm13.2$ & $0.117\pm0.002$ & $0.229\pm0.006$ & $0.484\pm0.012$ \\
Figure 2 & 200 GeV       & $116.3\pm3.0$ & $243.4\pm7.4$ & $597.4\pm18.3$ & $0.129\pm0.003$ & $0.271\pm0.009$ & $0.665\pm0.019$ \\
Cu-Cu 0--10\% & 62.4 GeV & $105.1\pm2.6$ & $210.1\pm6.3$ & $-$            & $0.117\pm0.003$ & $0.234\pm0.006$ & $-$ \\
Figure 3 & 200 GeV       &  $95.1\pm2.4$ & $236.3\pm7.0$ & $615.4\pm19.9$ & $0.106\pm0.002$ & $0.263\pm0.008$ & $0.685\pm0.020$ \\
$pp$     & 62.4 GeV      &  $90.3\pm2.3$ & $217.8\pm6.6$ & $-$            & $0.100\pm0.003$ & $0.242\pm0.007$ & $-$ \\
\hline
\end{tabular}%
\end{center}}
\end{table*}

From Figures 6 and 7, one can see that $T_{03}$ and
$\langle\beta_T\rangle_3$ decrease significantly, $T_{02}$ and
$\langle\beta_T\rangle_2$ decrease slightly, and $T_{01}$ and
$\langle\beta_T\rangle_1$ almost do not change with the decrease
of centrality from central to peripheral collisions. Meanwhile,
the relative values of $T_{03,02}$ and
$\langle\beta_T\rangle_{3,2}$ decrease significantly, and the
relative values of $T_{01}$ and $\langle\beta_T\rangle_1$ decrease
slightly. The tendencies of relative $T_{0i}$ and
$\langle\beta_T\rangle_i$ are the same as that of relative $T_i$
shown in Figure 5(b), but different from that of relative $V_i$
shown in Figure 5(a). The consistent results for relative $T_i$,
$T_{0i}$, and $\langle\beta_T\rangle_i$ are understandable,
because all three reflect $\langle p_T\rangle$. They are different
from relative $V_i$ which reflects the yields or normalization.

In higher energy collisions at the Large Hadron Collider, if the
range of $p_T$ spectra is wide enough, the multi-region structure
of $p_T$ spectra is usually observed~\cite{12-a,12-b,12-c}. The
multi-component standard distribution in the framework of
multi-source thermal model should describe the multi-region
structure naturally, where the number of components may be large
enough. Although lots of parameters are needed in the case of
multi-component, the physics explanation is acceptable due to the
multi-source of particles and multi-generations of strings in the
complex collisions. In lower energy collisions at a few GeV, if
the range of $p_T$ spectra is not too wide, a two-component
standard distribution is suitable, where the first component
describes the soft process and the second component describes the
hard process. If $T_1\approx T_2$, a sole component is
approximately applicable.

The interpretation of the results in terms of generations of
strings seems not satisfactory at very low collision energies,
though at which we may consider alternatively the formation of
nucleon clusters in the system, if the generations of strings are
not available. Meanwhile, the number of components may be 2 or
even 1, which is the special cases of the multi-source. In our
opinion, the multi-source thermal model can be used by
experimental collaborations to describe and extrapolate the data
on multi-particle productions at low and high energies. We believe
that the picture of multi-source is suitable for various
collisions, though the physics interpretations on sources
themselves may be different at different energies.

We would like to point out that although other functions such as
the Tsallis and \(q\)-dual distributions can fit the \(p_T\)
spectra by fewer parameters~\cite{16g,16h,16i,16j,52,51}, they
covered up the fine multi-region structure of the \(p_T\) spectra.
To study the volume, temperature, and flow velocity dynamics of
the system evolution in the framework of multi-source thermal
model~\cite{12,13,14,15}, we are inclined to the multi-component
standard distribution which is also the most classic distribution
and ``thermometer" in the ideal gas model in thermodynamics.
Usually, the standard-thermometer obtains higher temperature than
the Tsallis-thermometer~\cite{60}.

Using the multi-source thermal model to describe emission from the
different stages of the system evolution looks like an
approximation to a more dynamic one, like a well-know hydrodynamic
model~\cite{67,68,69}. However, in a hydrodynamic model, there is
typically one freeze-out temperature and a distribution of flow
velocities on the freeze-out surface (surface of constant
temperature) with which hadrons are produced. The present work
treats the system evolution at different stages by different
temperature and flow velocities.

Indeed, the approach used in the present work is different from
the widely used hydrodynamic one in some ways. According to the
hydrodynamic models, one freeze-out temperature can describe the
$p_T$ range only up to 3 GeV/$c$. Anything above that is typically
assumed to be non-thermal. However, in our opinion, the two types
of treatments may be consistent if one considers a multi-component
hydrodynamic model in which two or three sets of parameters may be
used. Of course, one may have different explanations for the
second and third components in the thermal and hydrodynamic
descriptions.

In addition, the present work just uses the thermal description
for particle productions based on the generations of strings in
general parton's interactions, but not the jet productions due to
head on parton's collisions. At the string or parton level, one
can see that the highest kinetic freeze-out temperature ($T_{03}$)
reaches up to above 700 MeV for high-$p_T$ ($\sim16$ GeV/$c$)
particle productions in central Au-Au collisions at the RHIC.
Although this temperature is very high and its contribution range
is very wide, its fraction is very low. Combining with low-$p_T$
particles, the temperature is only $\sim120$ MeV which is less
than the well-know chemical freeze-out temperature ($\sim150$--160
MeV~\cite{51-a,51-b,51-c,51-d}).

Before summarizing and conclusions, it should be noted that the
present work studies $V_i$ dynamics which is different from the
volume obtained using the correlations method~\cite{69a}. In the
latter, the pion source radii obtained from the two-pion
Bose-Einstein correlation functions in central Pb-Pb collisions at
2.76 TeV to be similar to the nuclear size. The present work
obtained the source volume to increase quickly from the parton
size ($V_3$) to the volume of an expanding fireball ($V_1$) which
is much larger than the nuclear size. As mentioned in the last
section, we have $k_i=V_i/\sum V_i$. This results in
$k_1>99.5\%-99.9\%$ from Table 1. The parameters from the first
component can represent the ones from the average weighted $k_i$
over the multiple components. In addition, $V_3$ is very small. To
show $V_3$ appropriately, $V_1$ and $V_2$ are showed by high
precise uncertainties so that three $V_i$ have the same number of
decimal places.
\\

{\section{Conclusions}}

In the framework of multi-source thermal model, we have studied
the fine multi-region structure of the transverse momentum spectra
of \(\pi^0\) produced in mid-rapidity region in Au-Au, Cu-Cu, and
\(pp\) collisions at the RHIC. We are inclined to the
multi-component standard distribution which can be used to extract
the parameters for the fine structure of the spectra since it is
the most classical distribution in the ideal gas model, though
other functions such as the Tsallis and \(q\)-dual distributions
can fit the spectra by fewer parameters. In this work, the
temperature and flow effect is separated by an alternative method
which is almost model-independent.

The significance of the present work is in the methodology, though
it is a simple application of the multi-component standard
distribution. The success of this work reflects that the classical
theory can still play a great role in the field of relativistic
collisions. In nucleus-nucleus collisions, the volume,
temperature, and flow velocity parameters decrease with the
decrease of centrality from central to peripheral collisions.
These parameters also decrease with the decrease of nuclear size
and collision energy. It is natural that at lower (higher) energy,
fewer (more) components of the distribution or generations of the
strings are needed.

Combining with the string model phenomenology, in the case of
using the three-component standard distribution, we could
conjecture that the high-temperature term corresponds to the
hadronization of the first generation strings, and the
intermediate- and low-temperature terms correspond to the
hadronization of the second and third generation strings
respectively. The third generation strings are from the breaking
of the second generation strings, and the second generation
strings are from the breaking of the first generation strings.

From the first generation strings to the third ones, the
increasing volume parameter and the decreasing temperature (flow
velocity) parameter describe the volume and temperature (flow
velocity) dynamics of the system evolution respectively. Obtained
$V_i$ and $T_i$, $i=1$ to 3 values, see Table 1, indicate a rapid
expansion of the system, which gradually cools down, causing the
particle velocity to decrease, see Table 2. Expecting the
formation of the first generation of strings to happen during the
initial stage of the relativistic collisions, the highest
temperature ($T_{03}$, see Table 2) could be attributed to the
initial temperature of the system at the string or parton level.
\\
\\
{\bf Acknowledgements} The authors would like to thank the
anonymous reviewer for his/her many constructive comments. The
work of Shanxi Group was supported by the National Natural Science
Foundation of China under Grant No. 12147215, the Shanxi
Provincial Natural Science Foundation under Grant No.
202103021224036, and the Fund for Shanxi ``1331 Project" Key
Subjects Construction. The work of K.K.O. was supported by the
Agency of Innovative Development under the Ministry of Higher
Education, Science and Innovations of the Republic of Uzbekistan
within the fundamental project No. F3-20200929146 on analysis of
open data on heavy-ion collisions at RHIC and LHC.
\\
\\
{\bf Data Availability Statement} The data used to support the
findings of this study are included within the article and are
cited at relevant places within the text as references.
\\
\\
{\bf Declarations}
\\
\\
{\bf Conflicts of Interest} The authors declare that there are no
conflicts of interest regarding the publication of this paper. The
funding agencies have no role in the design of the study; in the
collection, analysis, or interpretation of the data; in the
writing of the manuscript; or in the decision to publish the
results.
\\
\\

\end{multicols}
\end{document}